\documentclass[prl,superscriptaddress,twocolumn,floatfix,showpacs,amsmath,amssymb]{revtex4}

\usepackage{graphicx}
\usepackage{dcolumn}
\usepackage{bm}
\usepackage[latin1]{inputenc}

\begin{document}


\title{Microscopic Evidence of Spin State Order and Spin State Phase Separation in Layered Cobaltites RBaCo$_2$O$_{5.5}$ with R=Y, Tb, Dy, and Ho}

\author{H.~Luetkens}
\email{hubertus.luetkens@psi.ch}
\affiliation{Laboratory for Muon-Spin Spectroscopy, Paul Scherrer Institut,
CH-5232 Villigen PSI, Switzerland}
\author{M.Stingaciu}
\affiliation{Laboratory for Development and Methods, Paul Scherrer Institut, CH-5232 Villigen, Switzerland}
\author{Yu.G. Pashkevich}
\affiliation{A.A. Galkin Donetsk Phystech NASU, 83114 Donetsk, Ukraine}
\author{K. Conder}
\affiliation{Laboratory for Development and Methods, Paul Scherrer Institut, CH-5232 Villigen, Switzerland}
\author{E. Pomjakushina}
\affiliation{Laboratory for Development and Methods, Paul Scherrer Institut, CH-5232 Villigen, Switzerland}
\affiliation{Laboratory for Neutron Scattering, Paul Scherrer Institut \& ETH Z\"urich, CH-5232 Villigen, Switzerland}
\author{A.A. Gusev}
\affiliation{A.A. Galkin Donetsk Phystech NASU, 83114 Donetsk, Ukraine}
\author{K.V. Lamonova}
\affiliation{A.A. Galkin Donetsk Phystech NASU, 83114 Donetsk, Ukraine}
\author{P.~Lemmens}
\affiliation{IPKM, TU Braunschweig, D-38106 Braunschweig, Germany}
\author{H.-H.~Klauss}
\affiliation{IFP, TU Dresden, D-01069 Dresden, Germany}

\date{\today}

\begin{abstract}
We report muon spin relaxation measurements on the magnetic structures of
RBaCo$_2$O$_{5.5}$ with R=Y, Tb, Dy, and Ho. Three different phases, one ferrimagnetic and two antiferromagnetic, are identified below 300~K. They consist of different ordered spin state arrangements of high-, intermediate-, and low-spin Co$^{3+}$ of CoO$_6$ octahedra. Phase separation into well separated regions with different spin state order is observed in the antiferromagnetic phases. The unusual strongly anisotropic magnetoresistance and its onset at the FM-AFM phase boundary is explained.
\end{abstract}



\maketitle


Transition metal oxides exhibit a rich variety of interesting properties like spin, orbital
and charge order, giant magneto resistance (MR), and metal-insulator (MIT) transitions.
These properties reflect electronic correlations and an interplay of partly competing
degrees of freedom that may suppress long range ordered states. Of special interest
within this context are intrinsic and self-organized superstructures on microscopic
length scales. Such inhomogeneous states may be induced by doping as in cuprates or
manganites. However, chemical substitutions lead to considerable structural disorder
that may impede the understanding of intrinsic correlation effects. The layered
cobaltite RBaCo$_2$O$_{5.5}$ (R= rare earth) is a prominent example among strongly correlated electron systems, where ordered electronic structures and unconventional transport phenomena exist without extrinsic doping and having all Co ions in the trivalent
state \cite{Taskin05a,Moritomo00,Maignan99,Akahoshi99}. Thereby correlated
phases with a minimum of structural disorder can be studied.

RBaCo$_2$O$_{5.5}$ exhibits an orthorhombic crystal structure (\emph{Pmmm}), which is derived from the basic
perovskite by a doubling along the crystallographic $b$ and $c$ directions ($a_p \times
2a_p \times 2a_p$ unit cell, with $a_p$ being the cell parameter of the cubic
perovskite). The doubling of the $b$-axes originates from an alternation of CoO$_5$
square pyramids and CoO$_6$ octahedra along this direction, while the doubling along $c$
is due to the layer stacking of [BaO][CoO$_2$][RO$_{0.5}$][CoO$_2$] planes. In contrast
to LaCoO$_3$ the octahedra and pyramids are heavily distorted
\cite{Moritomo00,Plakhty05,Pomjakushina06}. These distortions support a variety of
Co$^{3+}$ spin states (low spin (LS, S=0), intermediate spin (IS, S=1), and high spin
(HS, S=2)) \cite{Wu01,Zhitlukhina07} as function of crystallographic environment and
temperature. A nonuniform spin state distribution - spin state order (SSO) due to a
complex interplay of electron-spin-orbital-lattice degrees of freedom has been
highlighted recently for RBaCo$_2$O$_{5.5}$
\cite{Coutanceau95,Foerster01,Khomskii04,Fauth02}. RBaCo$_2$O$_{5.5}$ shows a
series of phase transitions, namely a MIT below $T_\mathrm{MI}$ in the paramagnetic (PM)
phase, a PM to ferro(ferri)magnetic (FM) transition at $T_\mathrm{C}$, a FM to
antiferromagnetic (AFM1) at $T_\mathrm{N1}$ which is accompanied by the onset of strong
anisotropic magneto-resistive effects, and a AFM1 to antiferromagnetic (AFM2) phase
transition at $T_\mathrm{N2}$. Various contradicting magnetic structures including
different spin states of the Co$^{3+}$ ions and also SSO have been proposed, based on
neutron diffraction \cite{Fauth02,Soda03,Plakhty05,Frontera06}, macroscopic measurements
\cite{Taskin03}, and theoretical models \cite{Khalyavin05}, but no consensus has been reached.

In this Letter we report magnetic structures of four powder samples of
RBaCo$_2$O$_{5+\delta}$ with $\delta\approx 0.5$ and R = Y, Tb, Dy, and Ho determined by
means of muon spin relaxation ($\mu$SR). Our main result is that irrespective of the
rare earth ion a homogenous FM phase with ferrimagnetic SSO of IS and HS states develops
through two first order phase transitions into phase separated AFM1 and AFM2 phases with
different types of antiferromagnetic SSO. We argue, that the SSO and the phase
separation play a similar role as intrinsic inhomogeneities like doping do in cuprates and manganites. The specific
SSO in this cobaltite is also responsible for its unusual transport properties.

Powder samples of RBaCo$_2$O$_{5+\delta}$ were synthesized by conventional solid state reaction
techniques and subsequent oxygen content adjustment. Details of the sample preparation
and determination of the oxygen content $\delta$ with an accuracy of $\pm$0.01 can be found in Ref.~\cite{Conder05b}.

The $\mu$SR technique utilizes positively charged $\mu^+$ implanted at interstitial lattice sites that probe local magnetic fields via the $\mu^+$ spin precession frequency, which can be observed from homogenous phases that extend over at least tens to hundreds of unit cells. The amplitude
of the precession signal is proportional to the volume fraction of the corresponding magnetic phase. Therefore $\mu$SR is an ideal tool to investigate phase separation phenomena in magnetic materials, see e.g. \cite{Reotier97}.

\begin{figure}[htbp]
\center{\includegraphics[width=0.8\columnwidth,angle=0,clip]{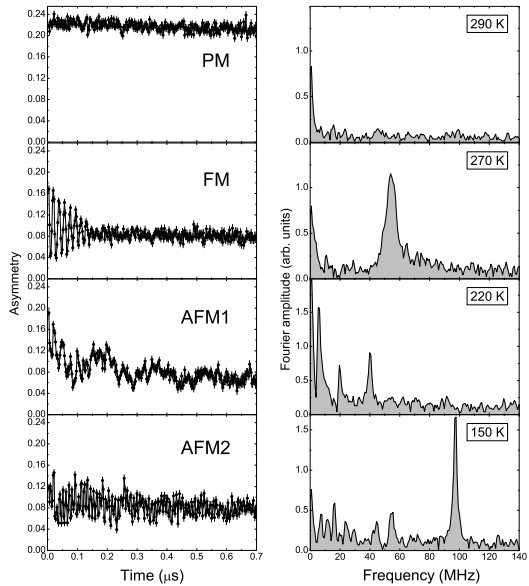}}
    \caption[]{Characteristic ZF-$\mu$SR time spectra (left panel) and corresponding Fourier spectra  (right panel) for temperatures within each of the different magnetic phases of YBaCo$_2$O$_{5.49}$.}
\label{spectra-and-fourier}
\end{figure}

In Fig.~\ref{spectra-and-fourier} zero field (ZF) $\mu$SR time spectra
and the corresponding Fourier analysis of YBaCo$_2$O$_{5.49}$ are shown for characteristic temperatures regimes.
Below $T_\mathrm{C}$ a well defined spontaneous $\mu^+$ spin precession develops indicative for the long-range magnetic order in the FM state.
Two different magnetic $\mu$SR signals are observed showing that two magnetically inequivalent interstitial lattice sites are occupied by the muons: one which is associated with the high oscillation frequency $\omega_\mathrm{HF}$ and another with a low frequency $\omega_\mathrm{LF}$ which is too strongly damped for the Y and Ho compounds, but measurable for Tb and Dy. The amplitude of the oscillating signal indicates that each of the two different sites is 50\% occupied.
Below $T_\mathrm{N1}$, the $\mu$SR spectra change drastically and several
superimposed oscillation frequencies are observed. Below $T_\mathrm{N2}$ the complexity of the $\mu$SR spectrum increases further indicating even more magnetically inequivalent $\mu^+$ sites. All data
have been fitted directly in the time domain by a superposition of exponentially damped oscillations and a practically undamped 1/3 fraction of the total observable asymmetry. The 2/3 oscillating and 1/3 undamped $\mu$SR signal fractions evidences the complete static magnetic ordering of the powder samples, since in a spatial average only 2/3 of the magnetic field components are perpendicular to the $\mu^+$ spin and cause a precession \cite{Reotier97}.

\begin{figure}[htbp]
\center{\includegraphics[width=0.9\columnwidth,angle=0,clip]{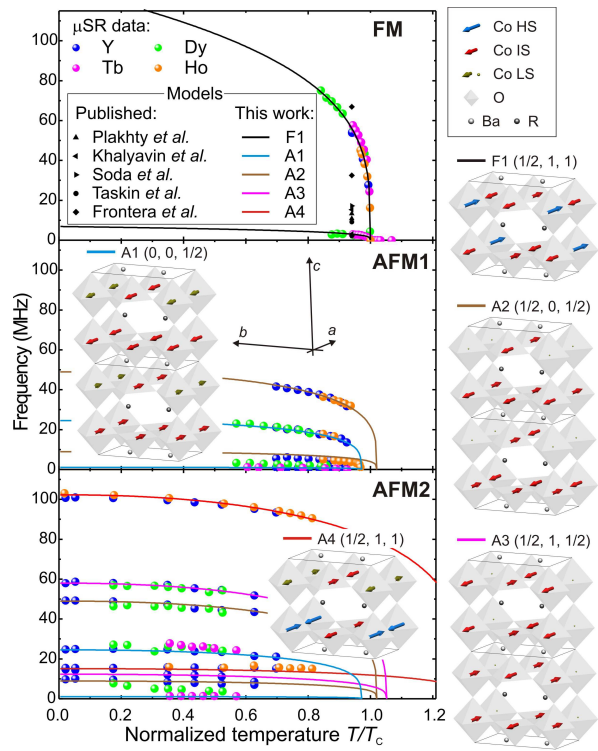}}
    \caption[]{Muon spin precession frequencies as a function of normalized temperature $T/T_\mathrm{C}$
    for RBaCo$_2$O$_{5+\delta}$ with R = Y, Tb, Dy, and Ho with $\delta\approx
    0.5$. The solid lines present the fit/calculation for Y
    for respective magnetic structures, see text. Pairs of lines which fit with the same power law and belong to the same structure are drawn in the same color. Magnetic reflections above magnetic structures are given for the unit cell $a_p \times 2a_p\times 2a_p$. For the FM phase the expected frequencies for various magnetic structures \cite{Soda03,Plakhty05,Frontera06,Taskin03,Khalyavin05} have been calculated for $T=0.94 T_\mathrm{C}$, see text.}
    \label{All-Separated}
\end{figure}

The measured frequencies are displayed in Fig.~\ref{All-Separated} for all samples as a function of normalized temperature
$T/T_\mathrm{C}$. For clarity, the frequencies for the different magnetic phases (FM, AFM1, and AFM2) have been drawn in separate
diagrams. The magnetic transition temperatures obtained from the $\mu$SR measurements are listed in Tab.~\ref{table}.
%
\begin{table}
\caption{Magnetic transition temperatures obtained by ZF-$\mu$SR for RBaCo$_2$O$_{5+\delta}$ with R = Y, Tb, Dy, and Ho.\label{table}}
\begin{ruledtabular}
\begin{tabular}{l c c c c}
& Y$^{3+}$ & Tb$^{3+}$ & Dy$^{3+}$ & Ho$^{3+}$ \\
$\delta$  &  0.49(1) &  0.50(1) & 0.50(1)  &  0.47(1)  \\
$T_\mathrm{C}$ (K) &  287(1) &  281(1) & 285.5(1.0) &  283(0.5) \\
$T_\mathrm{N1}$ (K) &  267(3) &  262(3) & 245(10)  &  273.5(0.5)  \\
$T_\mathrm{N2}$ (K) &  200(5) &  165(5) & 155(5)  &  235(5)  \\
\end{tabular}
\end{ruledtabular}
\end{table}
Qualitatively and quantitatively all samples show similar $\mu^+$SR frequencies indicating microscopically very similar magnetic structures, see Fig.~\ref{All-Separated}.
To evaluate quantitatively the consistency of the observed $\mu^+$SR frequencies with a specific magnetic structure model, the $\mu^+$ position
in the lattice has to be known. Therefore electronic potential calculations have been performed using a
modified Thomas - Fermi approach \cite{Reznik95}  and own structural data. This procedure has been verified on similar
oxide crystal structures where experimentally determined $\mu^+$ sites are available \cite{Holzschuh83}. In RBaCo$_2$O$_{5.5}$ two inequivalent $\mu^+$ sites are obtained from the
calculations which are located at N1=(0 0.36 0) and N2=(0.31 0 0) positions in the \emph{Pmmm} crystal
structure, i.e. about 1~$\AA$ away from apical oxygens in the BaO-plane. These are typical sites, which have also been
found e.g. in high-$T_\mathrm{C}$ cuprate superconductors \cite{Klauss04}.

For a given magnetic structure to be tested for consistency with the $\mu$SR
data, the local magnetic dipole fields at the two $\mu^+$ sites can be calculated numerically. Since there are
two $\mu^+$ sites in the lattice, two different $\mu$SR frequencies are calculated for a
homogenous phase. These two frequencies show a typical ratio which is characteristic for
the magnetic structure due to the symmetries of their corresponding $\mu^+$ sites. Additionally,
they have to display the same $T$ dependence. Finally, the magnetic model has to reproduce
the magnetic Bragg peaks that have been detected in neutron diffraction studies
\cite{Fauth02,Plakhty05,Soda06,Frontera06,Khalyavin07,Chernenkov07}, which strongly
reduces the number of structures to be tested.

The magnetic structure which is able to consistently describe the observed $\mu$SR
frequencies in FM phase is a checkerboard-like HS/IS AFM SSO on the octahedral sites
with IS AFM order on the pyramids and all moments pointing along the crystallographic
$a$-axis, see the F1 structure in Fig.~\ref{All-Separated}. The $T$ dependence
of the two frequencies observed in the FM phase can be perfectly described by the same power law $\omega(T)=\omega(0)(1-(T/\gamma T_\mathrm{C})^{\alpha})^{\beta}$ with $\omega_{HF}(0)$=127.7~MHz,
$\omega_{LF}(0)$=6.9~MHz, $\alpha = \gamma = 1$, and $\beta=0.290(5)$. This is reproduced by the dipole field
calculation by using the zero temperature HS and IS values $M^\mathrm{oct}_0$ as listed
in Tab.~\ref{table2} for the magnetic moments of Co in the two octahedral sites and
$M^\mathrm{pyr}_0=2\, \mu_\mathrm{B}$ (IS) on the pyramidal site. The calculation is
shown as a solid black line in Fig.~\ref{All-Separated}. The exponent $\beta=0.290(5)$
points to a 3D-Ising character of the interactions consistent with the observed
Ising-like anisotropy \cite{Taskin03,Khalyavin03,Zhou05}. The ferrimagnetic model F1 is
compatible with neutron scattering results
\cite{Fauth02,Plakhty05,Frontera06,Khalyavin07,Chernenkov07} since it conforms to the
most intense (1/2,1,1) magnetic reflex. The structural doubling along the $a$-axis and
the onset of \emph{Pmma} symmetry below MIT as observed in Gd \cite{Chernenkov05} by
x-ray and in Dy \cite{Chernenkov07} by neutron diffraction studies is also naturally
obtained by this magnetic structure, since the alternating IS and HS Co$^{3+}$ with
different ionic radii modulate the structure accordingly.

\begin{table}
 \caption{Magnetic moments $m^\mathrm{oct}(T)$=$M^\mathrm{oct}_0(1$-$(T/\gamma T_\mathrm{C})^\alpha)^\beta$ of the two different octahedral
Co for R = Y which are used for the calculation of the colored lines in Fig.~\ref{All-Separated}. In all phases, the pyramidal Co are in the IS state with $M^\mathrm{pyr}_0=2~\mu_\mathrm{B}$. The corresponding most intense magnetic Bragg reflex is also listed.\label{table2}}
 \begin{ruledtabular}
 \begin{tabular}{l c c c c c c}
Phase & Reflex & $M^\mathrm{oct}_0$/$\mu_\mathrm{B}$ & $M^\mathrm{oct}_0$/$\mu_\mathrm{B}$ & $\alpha$ & $\beta$ & $\gamma$ \\
 F1 & (1/2,1,1)  &  4.4 (HS) &  2.2 (IS) & 1.0  &  0.290(5) & 1 .00\\
 A1 & (0,0,1/2) &  1.7 (IS) &  0.3 (LS) & 2.2  &  0.26(1) & 0.97 \\
 A2 & (1/2,0,1/2) &  2 (IS) &  0 (LS) & 2.2 &  0.26(1) & 1.02 \\
 A3 & (1/2,1,1/2) &  2 (IS) &  0 (LS) & 2.2 &  0.29(1) & 1.05 \\
 A4 & (1/2,1,1) &  4.1 (HS) &  1.2 (LS) & 2.2  &  0.29(1) & 1.30 \\
 \end{tabular}
 \end{ruledtabular}
 \end{table}

Several previously proposed models for the FM phase have also been examined by
calculating the magnetic dipole fields at the two $\mu^+$ sites. We ensured that all
models give the same net magnetization (0.5~$\mu_\mathrm{B}$/F.U. for single crystal,
i.e. approximately 0.25~$\mu_\mathrm{B}$/F.U. for powders at $T=0.94 T_\mathrm{C}$,  by
scaling the local moments for the magnetic models
\cite{Soda03,Plakhty05,Frontera06,Taskin03,Khalyavin05}. None of the published models is
close to the observed $\mu$SR frequencies, see Fig.~\ref{All-Separated}. In particular,
none of the models exhibits the correct ratio of the two fields at the two $\mu^+$ sites, which is
independent on the scaling.

In the AFM phases more than two $\mu$SR frequencies are observed. Yet, these
frequencies always appear in pairs of one HF and one LF signal whose $T$ dependence can
be fitted with the same power law, while the other pairs follow different laws. Each
pair of fitted lines is shown with a unique color in Fig.~\ref{All-Separated} (only the
fit for the Y compound is shown). Therefore, we conclude that every pair of $\mu$SR
signals belongs to a separate magnetic phase with a different $T$ dependence of
its order parameter and different extrapolated N\'eel temperatures
($T^*_\mathrm{N}$=$\gamma T_\mathrm{C}$). A clear $\mu$SR precession with a small spread of dipole fields is obtained only if the volume of a homogenous phase extends over at least tens to hundreds of unit cells. Thus, we conclude that the first order FM-AFM1
phase transition in layered cobaltites occurs with a phase separation, i.e. with
simultaneous appearance of phases with different magnetic structures and different types
of SSO.

Depending on R, we deduce two types of SSO (A1 and A2) in the AFM1 phase and up to four different types of SSO (A1-A4) in the AFM2 phase, which appear at the respective phase transitions and develop on the cost the high temperature phases when the temperature is lowered.
The magnetic structures for all phases of the Y compound are shown in Fig.~\ref{All-Separated}. Their deduced sublattice magnetic moments and $T$ dependence are listed in Table~\ref{table2}. A common
feature of all structures is the AFM coupling of IS states on pyramidal sites along the
$c$-direction. The stability of pyramidal IS states in RBaCo$_2$O$_{5.5}$ has
recently been revealed by ab-initio calculations \cite{Pardo06}. Note that the
structures A1-A3 present just a different topology of the IS/LS pair distribution on
neighboring octahedral sites which have very similar self-energy. Probably, this is the
reason of the phase separation at the FM-AFM1 phase transition. The representative
magnetic reflexes of the structures F1 and A1-A4 are given in Tab.~\ref{table2}. All have been observed by neutron diffraction
\cite{Khalyavin07,Chernenkov07,Plakhty05,Soda06} in the corresponding phases, but were
not or differently assigned. Only the phase A3 has also been deduced from one neutron
study \cite{Fauth02}. Phase separation can be identified by a volume sensitive local probe like $\mu$SR or NMR. The observed low temperature AFM/SSO phase separation with IS
states on pyramids and with IS and LS (A1-A3) and HS (A4) states on octahedra is
consistent with the low temperature NMR study of the Y compound \cite{Itoh03}, where
four different types of Co species have been detected.

Now, we discuss some consequences of the observed magnetic structures. Antiferromagnetic
coupling in pyramids along c-axis is the main motive of all structures (F1 and A1-A4).
This is in accordance with Goodenough-Kanamori-Anderson (GKA) rules for IS
($t^5_{2g}e_g^1$) states on pyramids with the $e_g$-electrons occupying either the
$3z^2-r^2$ or the $x^2-y^2$ orbitals. Furthermore, one can show that all magnetic bonds
in the A1-A4 structures fulfill GKA rules (assuming a certain occupancy of the
orbitals). On the other hand GKA rules can not be satisfied for the F1 HS/IS  SSO
structure realized in the FM state, where one of the pyramid-octahedra magnetic bonds
remains frustrated in the G-type order. This frustration might explain the narrow
temperature range in which FM order is observed.

Finally, the detected magnetic structures are able to explain qualitatively the
transport properties of RBaCo$_2$O$_{5.5}$ in a localized picture of charge carriers
motion \cite{Maignan04}. The A1 phase consists of ferromagnetic $ab$-planes of Co$^{3+}$
in IS states along which "electron" (HS Co$^{2+}$ species, $t^5_{2g}e^2_g$) hopping is
allowed, while it is impossible along $c$-axis for all A1-A4 phase due to the so-called
spin blockade mechanism \cite{Maignan04,Taskin05b}. Less pronounced "hole" (LS Co$^{4+}$
species, $t^5_{2g}$) hopping is allowed through the channels of LS states along $a$-axis
in the A1 and A2 phases and along $c$-axis in the A3 phase. This explains the one order
of magnitude different resistivities along $c$- and $a$-axis in detwinned single
crystals of EuBaCo$_2$O$_{5.5}$ \cite{Zhou04b}. The phase separation into less and more
conductive phases in RBaCo$_2$O$_{5.5}$ leads to a resistivity behavior similar to low
doped manganites where conductive and isolating regions coexist. Also the observed
strong anisotropy of the MR \cite{Zhou05} can be understood on the basis of the deduced
magnetic structures. Due to the Ising like anisotropy a magnetic field along the
$a$-axis may switch the weakly AFM coupled neighboring FM planes in the A1 phase to the
FM alignment (as in usual metamagnetic antiferromagnets) making "electron" transport
along $c$ possible. As a result, a very large MR for fields along $a$ is observed.
Furthermore, the hopping process for both kind of charge carriers is strictly prohibited
in the F1 phase, even when the magnetic structure is canted by a magnetic field, which
explains the unusual onset of the MR phenomenon below $T_\mathrm{N1}$.

In conclusion, we present the first local probe ($\mu$SR) investigation on the magnetic
structures of the RBaCo$_2$O$_{5+\delta}$ system with $\delta\approx 0.5$. SSO is established in all magnetic phases. Phase separation into well separated regions with different SSO is observed in the AFM phases.
The deduced SSO magnetic structures are consistent with the magnitude of the 
magnetoresistance, its unusual anisotropy and its onset at the FM-AFM phase 
boundary.

\begin{acknowledgments}
This work was performed at the Swiss Muon Source, Paul Scherrer Institut, Villigen,
Switzerland. We acknowledge support by DFG, ESF-HFM and the NCCR MaNEP project. 
\end{acknowledgments}


\end{document}